\documentclass[12pt]{article}
\usepackage{graphicx}
\begin{document}

\pagestyle{empty}

\noindent
{\bf Studies of the Long Secondary Periods in Pulsating Red Giants. II. Lower-Luminosity Stars}

\bigskip

\noindent
{\bf John R. Percy and Henry Wai-Hin Leung\\Department of Astronomy and Astrophysics, and\\Dunlap Institute of Astronomy and Astrophysics\\University of Toronto\\Toronto ON\\Canada M5S 3H4}

\bigskip

{\bf Abstract}  We have used AAVSO visual and photoelectric V data, and the
AAVSO time-series package VSTAR and the Lomb-Scargle time-series algorithm
to determine improved pulsation periods, ``long secondary periods" (LSPs), and their amplitudes in 51 
shorter-period pulsating red giants in the AAVSO photoelectic photometry
program, and in the AAVSO LPV (long period variable) binocular program.  As is well known, radial
pulsation becomes detectable in red giants at about spectral type M0, with periods of about
20 days.  We find that the LSP phenomenon is also first detectable at about M0.
Pulsation and LSP amplitudes increase from near zero to about 0.1 at periods
of 100 days.  At longer periods, the pulsation amplitudes continue to increase, but 
the LSP amplitudes are generally between 0.1 and 0.2 on average. 
The ratios of LSP to pulsation
period cluster around 5 and 10, presumably depending on whether the pulsation
period is the fundamental or first overtone.  The pulsation and LSP phase
curves are generally close to sinusoidal, except when the amplitude is small, in which
case they may be distorted by observational scatter or, in the case of the
LSP amplitude, by the pulsational variability.  As with longer-period
stars, the LSP amplitude increases and decreases by a factor of two or more, for unknown reasons, on a median time
scale of about 20 LSPs.  The LSP phenomenon is thus present and similar in 
radially pulsating red giants of all periods.  Its cause remains unknown.

\medskip

\noindent
AAVSO keywords = AAVSO International Database; photometry, visual; pulsating variables; giants, red; period analysis; amplitude analysis

\medskip

\noindent
ADS keywords = stars; stars: late-type; techniques: photometric; methods: statistical; stars: variable; stars: oscillations

\medskip

\noindent
{\bf 1. Introduction}

\smallskip

In a previous paper (Percy and Deibert 2016), which is one of a series of our
papers about pulsating red giants, we addressed the question of the nature
and cause of the ``long secondary periods" (LSPs) which occur in about a third
of these stars, and whose cause is unknown.  In the present paper, we look especially at shorter-period
pulsating red giants in two samples: (1) stars in the AAVSO photoelectric (PEP) program,
which have both visual and PEP data; (2) shorter-period stars in the AAVSO
LPV Binocular Program (www.aavso.org/lpv-section-file-downloads).  These
data, though not as precise as, for instance, MACHO and OGLE data, have
the advantage that they have been sustained over many decades.

AAVSO PEP observations of pulsating red giants have already been analyzed
by Percy {\it et al.} (1996).  Now, there are an additional two decades of
data.  Robotic telescope PEP observations of similar
stars were analyzed by Percy {\it et al.} (2001), and merged AAVSO and robotic 
observations were analyzed
by Percy {\it et al.} (2008).

We hope to address scientific questions such as whether the LSPs occur in shorter-period,
lower-luminosity red giants, and whether their amplitudes, and their ratio to
the fundamental pulsation period are the same as in longer-period stars.
Shorter-period stars have smaller pulsation and LSP amplitudes but, for the same length
of dataset, yield more accurate values of the LSP,
and the timescale of its amplitude variation.  Since the pulsation periods are much less
than a year, they are slightly less likely to be complicated by one-cycle-per-year
aliases.  Ultimately, we would like to make progress in identifying the
nature and cause of the LSP phenomenon.
 
\medskip

\noindent
{\bf 2. Data and Analysis}

\smallskip

We used visual and PEP V observations from the AAVSO International Database (AID:
Kafka 2016), and the AAVSO VSTAR time-series analysis package (Benn 2013),
which includes both a Fourier and a wavelet analysis routine. 
Co-author HL was interested in comparing the results of VSTAR with those
from the Lomb-Scargle algorithm (implemented here with the astropy.stats.LombScargle routine within {\it Python} (www.python.org)) so we used
that also for some of the analysis.   Figure 1 shows the period spectrum
of the visual data on T Cen, obtained with the Lomb-Scargle algorithm. 
The best period, the one-year aliases, and the harmonics are present,
and marked.

\begin{figure}
\begin{center}
\includegraphics[height=7cm]{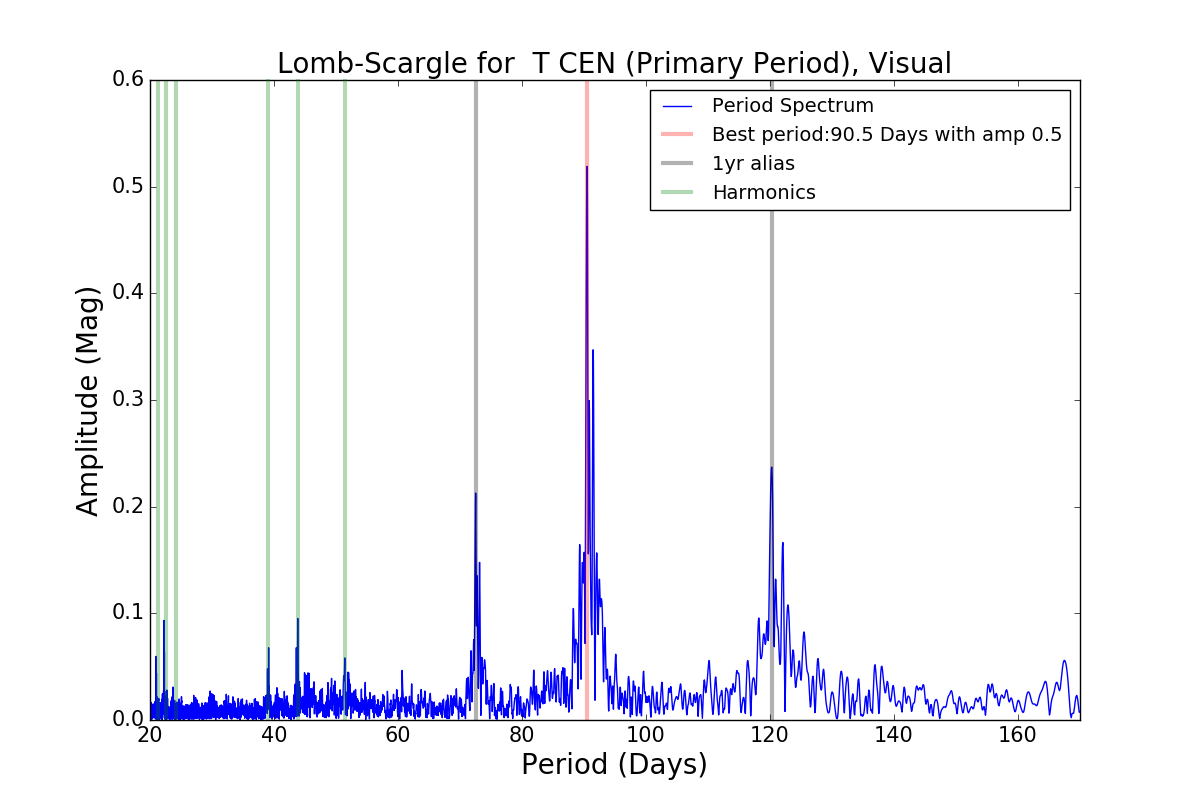}
\end{center}
\caption{The Lomb-Scargle period spectrum for T Cen visual data, showing
the best period (90.5 days), the one-year aliases, and the harmonics.}
\end{figure}

Our two samples of stars have some selection effects.  Those on the PEP
program were pulsating red giants which were on the AAVSO visual program in the early 1980's, but
had small amplitudes, and would therefore benefit from PEP observations.
The stars on the binocular program were presumably stars which were
reasonably bright, and had moderate to high amplitudes.  We chose to
analyze stars, from this program, with shorter periods, since we were especially interested in the LSP
phenomenon in such stars.

\medskip

\noindent
{\bf 3 Results}

\smallskip

\noindent
3.1 Periods and Amplitudes

\smallskip

We have determined improved periods and amplitudes in these 43 stars.
Table 1 lists stars in the AAVSO LPV Binocular Program with periods less
than about 120 days, for which there were sufficient data for analysis, along
with our results.  The columns give: the star name, the pulsation period PP,
the LSP, the ratio of LSP to PP, and the pulsation and LSP amplitudes.  In this and Table 2, some LSPs are close
to one year and, if their amplitudes are small, there is some possibility
that they are spurious (Percy 2015).
Table 2 lists stars on the AAVSO PEP program, for which there
were sufficient visual and PEP V data to determine periods (see Notes on Individual
Stars), along with our results.  The columns give: the star name, the
pulsation period PP, the LSP, the ratio of LSP to PP, and the
pulsation and LSP amplitudes.  In deciding on the value of the amplitude, we have usually given
greater weight to the photoelectric V data.  In any case, the PP and LSP
amplitudes of these stars vary with time (Percy and Abachi 2013).
Note that, although we concentrated on stars with shorter periods, there
were stars on the PEP program (RS Cnc, $\eta$ Gem, and SW Vir) which
had longer periods.  On the binocular program, X Her was listed as having
a period of 102 days, but was found to have a longer one; Y Lyn was listed
as having a period of 110 days, but was also found to have a longer one.

There were six stars (TV Psc, Z Eri, RR Eri, TV UMa, FP Vir and V1070 Cyg)
which were common to both programs.  The two authors decided to analyze them
independently.  For Z Eri, this yielded two different results for the
pulsation period -- 78.5 and 118.4 days.  For this star, the Fourier
spectrum yielded {\it several} peaks of comparable height, including those mentioned.  In the V data,
the highest was 239-243 days.  We conclude that, with the present data,
the pulsation period is indeterminate.  For RR Eri, we obtained two different
results for the LSP -- 366 and 742 days.  In the Fourier spectrum of the V
data, 742 days is marginally higher, whereas 366 days could well be a spurious
period.  Aside from these marginal differences, VSTAR and Lomb-Scarge gave
equivalent results.

\medskip

\noindent
3.2 Long Secondary Periods

\smallskip

For the stars in Table 1, Figure 2 plots the ratio of LSP to pulsation period LSP/PP against PP. 
The shaded area is a histogram projected on the y-axis,
with the scale on the top.  Percy and Deibert (2016) found the ratio LSP/PP to be about 5 when the pulsation
period was the fundamental period, and about 10 when the pulsation period was
the first overtone.  Figure 2 also shows clustering at about 5-6 and 8-10 and
also some stars with a ratio of 12-14.  The latter have short pulsation
periods, and may be pulsating in the second overtone.  This is consistent
with previous studies of pulsation modes in short-period, small-amplitude
red giants, which showed that many of these pulsate in the first or second overtone (e.g.
Percy and Bakos 2003).

\begin{figure}
\begin{center}
\includegraphics[height=7cm]{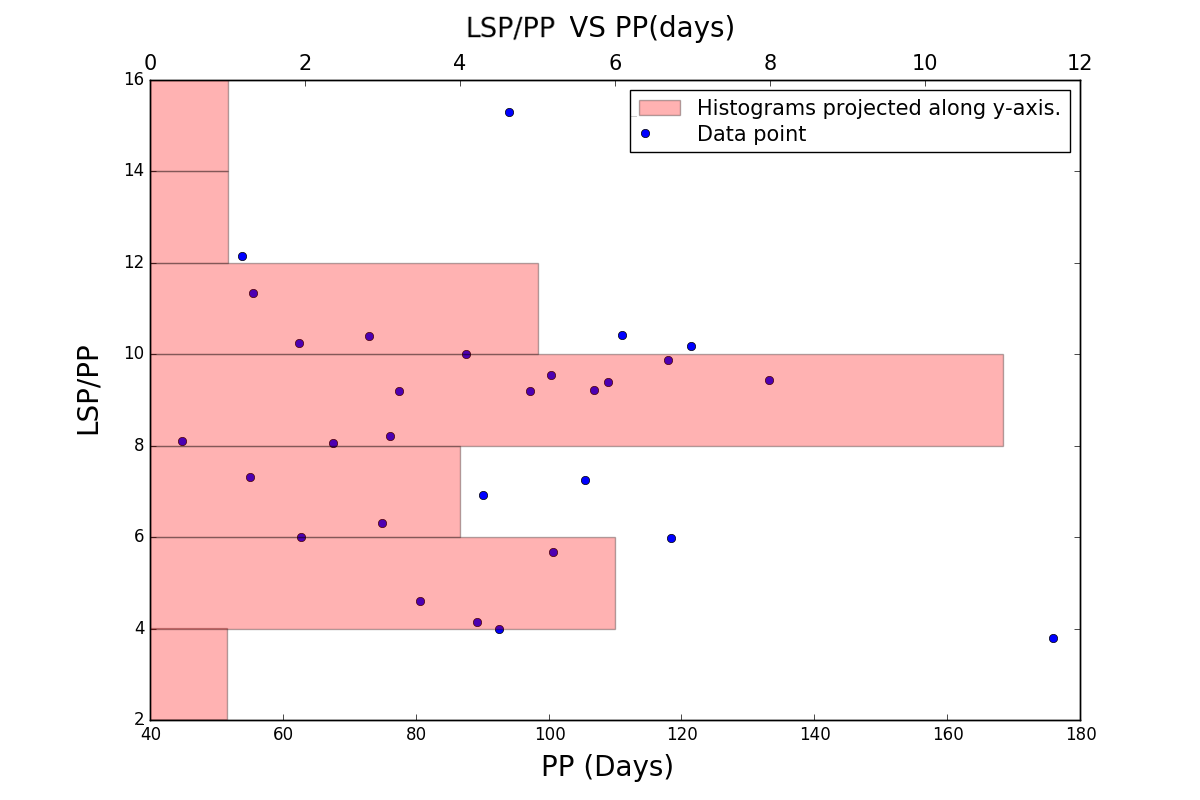}
\end{center}
\caption{The ratio of LSP to pulsation period PP, as a function of PP, 
for the stars in Table 1.  The shaded area represents a histogram of the data, with the
scale at the top.  The results are consistent with those of Percy and
Deibert (2016), namely that the values cluster about 5 and 10, depending
on whether the pulsation period is the fundamental period or the first
overtone.}
\end{figure}

\smallskip

\noindent
3.3. Amplitudes

\smallskip

Figure 3 plots the pulsation amplitude against the pulsation period.  
As is well-known (see Percy and Guler (1999) and the review by Kiss and Percy (2012), for instance),
pulsation sets in at about M0 spectral type, which corresponds to
periods of about 20 days.  The pulsation amplitude then increases with period, as
seen in Figure 4.

\begin{figure}
\begin{center}
\includegraphics[height=7cm]{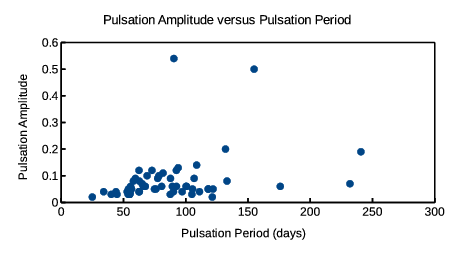}
\end{center}
\caption{The pulsation amplitude versus pulsation period.
As is well known from previous studies (reviewed by Kiss and Percy 2012), the shortest-period stars have
the smallest amplitudes; radial pulsational instability sets
in at spectral type approximately M0III.}
\end{figure}

\smallskip

Figure 4 plots the PP amplitude against the LSP amplitude.
The two are approximately equal for periods less
than about 100 days.  For longer periods, the LSP amplitude is
typically 0.05-0.2 (Percy and Deibert 2016).  The LSP phenomenon thus
becomes detectable at the same spectral type as radial pulsation.

\begin{figure}
\begin{center}
\includegraphics[height=7cm]{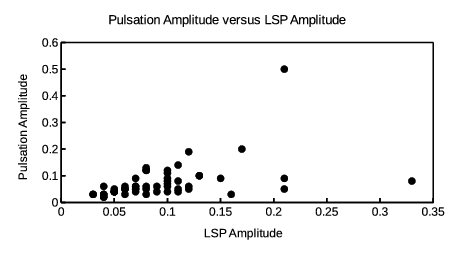}
\end{center}
\caption{The PP amplitude as a function of LSP amplitude, for
the stars in Table 2.  For longer-period stars, Percy and Deibert (2016) found
that the LSP amplitude was typically 0.2.  For the shorter-period stars,
both the LSP amplitude and the pulsation amplitude are approximately equal.
Both radial pulsation, and the LSP phenomenon set in at spectral type 
about M0III.}
\end{figure}

\medskip

\noindent
3.4 Phase Curves

\smallskip

For the stars in Table 1, the shape of the pulsation phase curve, and the
LSP phase curve were investigated by fitting them with a fifth-degree polynomial,
and comparing the result to a sine curve using the scipy.stats.pearsonr module in
{\it Python}.  The results are shown in figures 5 and 6.
For the pulsation phase curves,
there is a small dissimilarity for stars with amplitudes less than 0.1.  The
simplest explanation is that, for these, the shape of the true (sinusoidal?) phase curve is being
distorted by observational scatter.  The LSP phase curves also tend to be
less sinusoidal if the amplitude is less than 0.1; again, the simplest
explanation is that this is due to the distorting effect of observational scatter, and of the pulsational variability.
For the following stars, the LSP phase curve was flagged as being non-sinusoidal: 
V1070 Cyg, Z Eri, RX Lep, and XY Lyr.  Only two of the four have amplitudes
less than 0.10.   Percy and Deibert (2016) found,
from visual inspection, that the LSP phase curve of Y Lyn was clearly sawtooth,
rather than sinusoidal; here, we find it to be sinusoidal.

\begin{figure}
\begin{center}
\includegraphics[height=7cm]{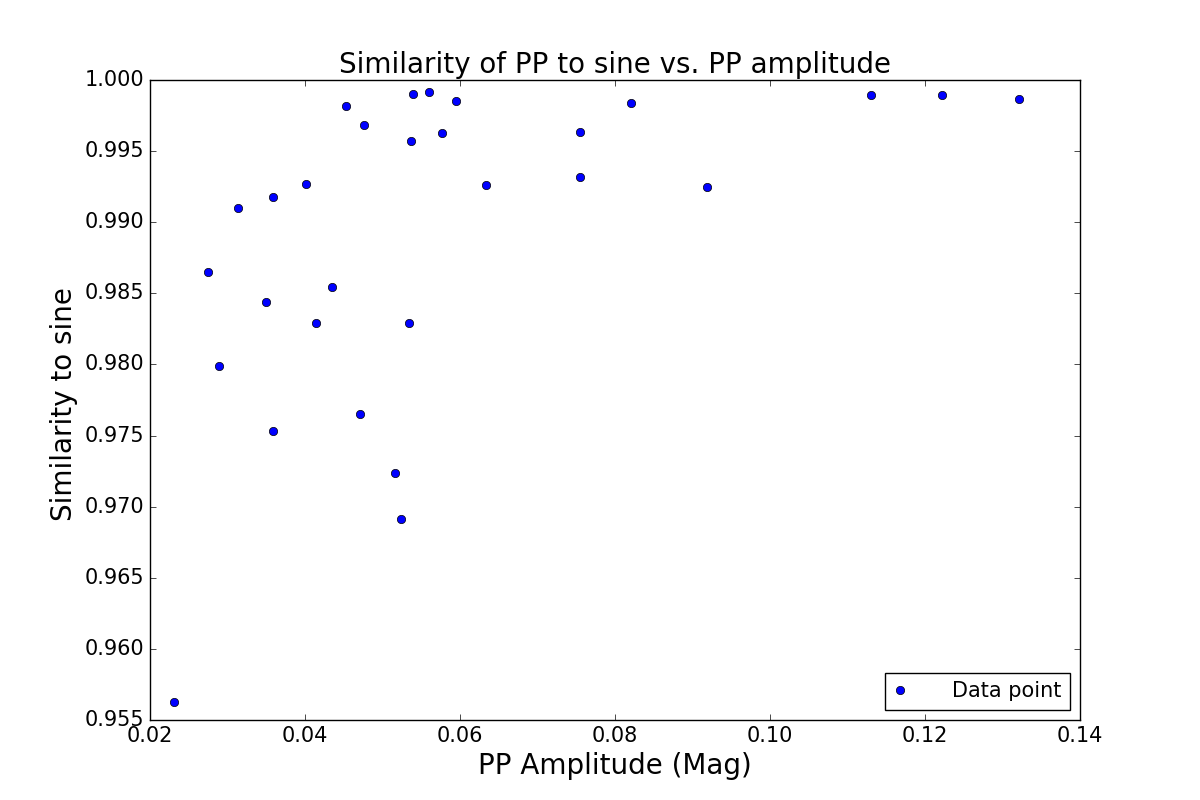}
\end{center}
\caption{The similarity of the pulsation phase curve to a sine curve,
as a function of pulsation amplitude.  Stars with small pulsation
amplitudes may have phase curves which differ slightly from sine curves,
perhaps because observational scatter distorts the phase curve.}
\end{figure}

\begin{figure}
\begin{center}
\includegraphics[height=7cm]{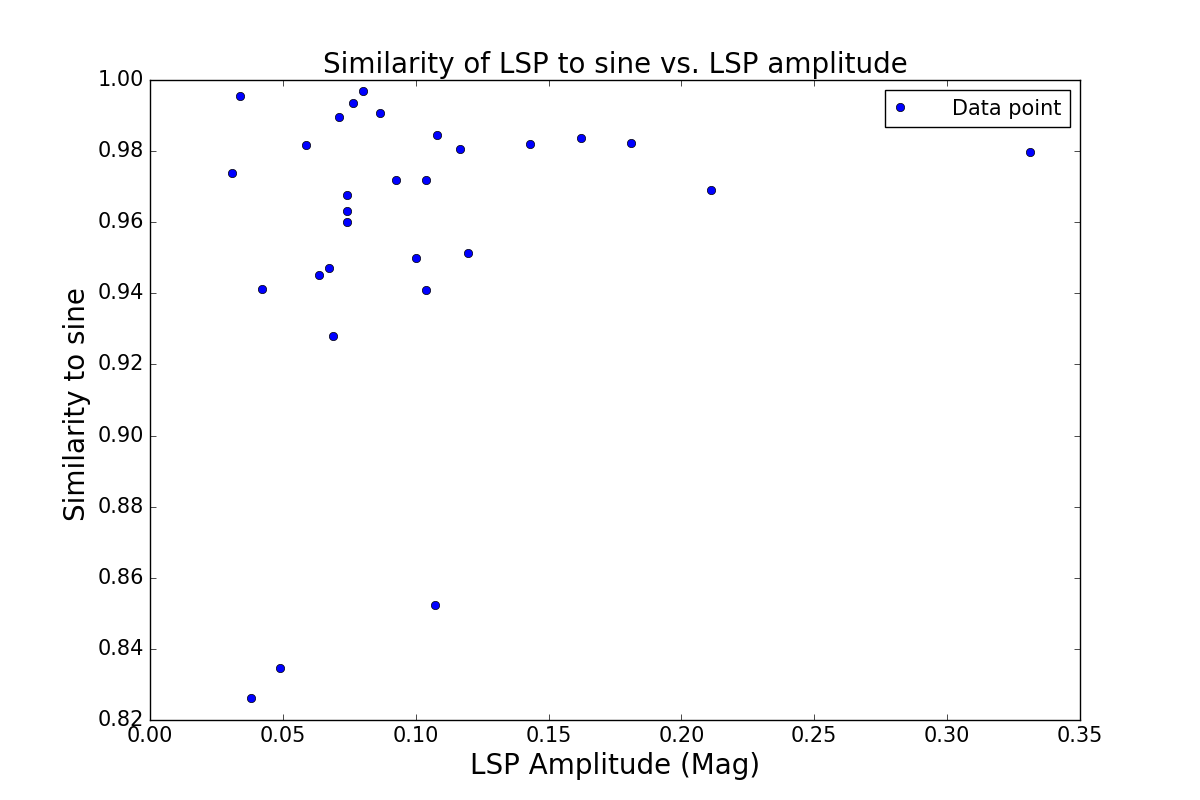}
\end{center}
\caption{The similarity of the LSP phase curve to a sine curve, as a
function of the LSP amplitude.  Stars with smaller LSP amplitudes may
have LSP phase curves which differ slightly from sine curves, perhaps
because observational scatter, and the pulsational variability distort the LSP phase curve.}
\end{figure}

\medskip

\noindent
3.5 Amplitude Variations

\smallskip 

Table 3 lists stars, from Table 1, which had sufficient data to investigate
variations in the LSP amplitude using the wavelet routine in VSTAR.  The columns list: the name of the star, the
range of the LSP amplitude, the number of cycles of LSP amplitude increase
and decrease N, and the ratio of the length L of these cycles to the LSP, where
L is the length of the dataset divided by N.
See our previous papers, especially Percy and Abachi (2013), for more discussion of the determination of these.
In particular: note that N and L are very approximate, because there is often
less than one cycle of increase and decrease in the dataset -- even though the dataset may be many decades long.

The median ratio of cycle length L to LSP is 21, slightly lower than the value 30
found by Percy and Abachi (2013); the difference is probably not significant. 
The LSP amplitudes vary by typically a factor of two, again in agreement
with the values found by Percy and Abachi (2013). 
Combining our results with those of Percy and Abachi (2013), there is no
obvious correlation between L/LSP and pulsation period i.e. with the radius of the
star.

\medskip

\noindent
{\bf 3.6 Notes on Individual Stars}

\smallskip

\smallskip

{\it $\chi$ Aqr:}  The V data give periods of 40.2 and 229 days, but the
amplitudes are small.  The visual data do not give reliable periods.

\smallskip

{\it RZ Ari:} The visual and V data give the same pulsation period (56.5 days)
but the LSP (507 days) is present in the V data only.

\smallskip

{\it W Boo:}  This star switches modes between 25 and 50 days (Percy and Desjardins 1996).
The present V data give a strongest period of 25.7 days.  The visual and V data both
give an LSP of 358-365 days, with small amplitude; it is suspiciously close
to one year, and may be spurious.

\smallskip

{\it VZ Cam:} There are no significant peaks in the visual or V data.

\smallskip

{\it RS Cnc:} The pulsation period of 240.8 days is present in both the
visual and V data.  There is a possible LSP of 2050 $\pm$ 100 days.

\smallskip

{\it DM Cep:} No periods could be found.

\smallskip

{\it FZ Cep:}  The V data give periods of 81.8 and 743 days.  The visual
data are inconclusive.

\smallskip

{\it T Cet:}  The pulsation period may be 161 or 288 days; these are aliases.
The LSP may be 1908 days, but this too may be an alias.

\smallskip

{\it FS Com:}  The visual and V data both give a pulsation period of 55.7 days.
The V data give an LSP of 689 days, in agreement with the result of Percy
{\it et al.} (2008).  The visual data do not show an LSP.

\smallskip

{\it GK Com:} The results are uncertain.

\smallskip

{\it W Cyg:}  The visual data give periods of 132 and 259 days, as do the
V data.  We assume these to be the fundamental and first overtone periods.
There is no evidence for an LSP.

\smallskip

{\it AB Cyg:}  The V data give periods of 69.3 and 521 days.  The visual
data give an LSP of 529 days, but the pulsation period is uncertain.

\smallskip

{\it V973 Cyg:}  The visual data give periods of 40.5 and 362 or 394 days.
The V data give periods of 36.5 and 391 days.  In each case, the amplitude is small.

\smallskip

{\it V1070 Cyg:}  The visual data give periods of 62.8 and 639 days.  The
V data are inconclusive.

\smallskip

{\it V1339 Cyg:}  The visual and V data both give pulsation periods of
34.06 days.  The visual data give an LSP of 339 days, but the amplitude
is only 0.02, so it is uncertain.

\smallskip

{\it EU Del:}  The visual data give periods of 62.55 and 623.8 days.  The V data
give periods of 62.52 and 629.3 days.  This star is a ``prototype" of small-amplitude pulsating red giants.

\smallskip

{\it VW Dra:}  The data are noisy.

\smallskip

{\it AT Dra:}  Data are plentiful, but somewhat noisy.  There are no
significant peaks.

\smallskip

{\it AZ Dra:}  The visual data give a pulsation period of 44.4 days.  The
visual and V data both give an LSP of 357-360 days, with amplitudes about
0.10.  The period may be spurious.

\smallskip

{\it Z Eri:}  Both the visual and V data give periods of 78.5 and 730 days.

\smallskip

{\it RR Eri:} The V data give periods of 92.5 and 742 days.  The former is
weakly present in the visual data, but not the latter.  The GCVS period
is 97 days.

\smallskip

{\it $\eta$ Gem:} The pulsation period (232 days) is present in both the
visual and V data, but there is no obvious LSP.

\smallskip

{\it IS Gem:} Non-variable.

\smallskip

{\it X Her:} {\it The General Catalogue of Variable Stars} gives a period
of 95 days; we find 176 days.

\smallskip

{\it ST Her:}  Both the visual and V data give periods of 151 or 257 days.
These are aliases, and we cannot choose between them.

\smallskip

{\it AK Hya:} The data are noisy; there are no convincing peaks.

\smallskip

{\it IN Hya:}  The V data suggest a pulsation period of 87.9 days and an LSP
of 693 days.  The visual data are less clear, but may indicate a pulsation
period of 85.8 days.

\smallskip

{\it SS Lep:} Probably non-variable.

\smallskip

{\it Y Lyn:}  {\it The General Catalogue of Variable Stars} gives a period
of 110 days; we get 133.2 days.

\smallskip

{\it R Lyr:}  The visual and V data give a period of 53 $\pm$ 10 days.  There
is no LSP (Percy {\it et al.} 2008).

\smallskip

{\it V614 Mon:} The pulsation period is unclear, and the LSP in both the
visual and V data is dangerously close to one year.

\smallskip

{\it V533 Oph:}  The visual data give periods of 55.3 and 398 days.  The V
data give periods of 60.0 and 405 days.

\smallskip

{\it $\rho$ Per:} The visual and V data give the same pulsation period
(55 $\pm$ 0.5 days) but the LSP (723 days) is visible in the V data only.

\smallskip

{\it TV Psc:} The periods from the visual (55.1 days) and V data (55.0 days)
are consistent, even though the amplitudes are small -- 0.06 and 0.02 respectively -- but the LSP is uncertain: 403 days from the visual data, and 546 days
from the V data, both seemingly well-determined.

\smallskip

{\it TX Psc:}  No reliable periods could be found.

\smallskip

{\it XZ Psc:}  No reliable periods could be found.

\smallskip

{\it V449 Sco:}  There are no significant peaks.

\smallskip

{\it CE Tau:} The pulsation period (105 $\pm$ 2 days) and the LSP (1280 $\pm$ 10 days) are present in both the visual and V data.

\smallskip

{\it TV UMa:} The visual data give periods of 53.8 and 656 days; the V data
give periods of 50-60 and 627 days.

\smallskip

{\it VY UMa:}  The visual data give periods of 121.8 and 1200 days.
The V data are less certain, but suggest periods of 125 and 1160 days.

\smallskip

{\it VW UMa:}  The visual data give periods of 65.9 and 624 days; the
V data give periods of 66.2 and 618 days.

\smallskip

{\it SW Vir:}  The visual and V data give pulsation periods of 155.3 and 154
days, respectively.  The LSP is uncertain, but may be 1647 days.

\smallskip

{\it EV Vir:}  The data are too sparse to yield a result.

\smallskip

{\it FH Vir:}  The V data give a pulsation period of 59.6 days, in reasonable
agreement with the GCVS period of 70 days.  The LSP may be about 350-370 days,
but this is dangerously close to a year.

\smallskip

{\it FP Vir:}  The visual data give periods of 62.8 and 383 days; the V data
give periods of 67.0 and 369 days.

\medskip

\noindent
{\bf 4. Discussion}

\smallskip

We must emphasize the challenges and consequent uncertainties of our
work: the amplitudes of our stars are small; the data are primarily
visual, so they have limited accuracy and the possibility of spurious
signals.  Despite the decades-long database, some of the phenomena that we
are studying would benefit from an even longer one.

Our results should provide additional constraints on the cause of
the LSP phenomenon.  That cause must be able to act in lower-luminosity stars
as well as higher-luminosity ones.  At the same time, its amplitude must
approach zero in the lowest-luminosity pulsators, as the pulsation
amplitude does.  And whatever causes the cyclic changes in LSP {\it amplitudes}
must operate over the entire period/luminosity range.  Wood (2015) also
shows (his Figure 1) that LSPs extend to the lowest luminosities -- though
sparsely -- in the Large Magellanic Cloud.  It also appears that LSPs
occur on both the red giant branch and on the asymptotic giant branch.
The LSP mechanism must explain the tight correlation between the LSP
and the fundamental pulsation period, and the consistency of the LSP
amplitude variations over the entire LSP range.  It must also explain the
large prevalence of LSPs -- over 30 percent at all luminosities.  It cannot
be dependant on some rare process or configuration in the star.

Several possible causes of LSPs have been suggested in the literature e.g. Nicholls {\it et al.} (2009).
Some do not match all of the observations.  Others are inconsistent with
theoretical predictions, in their present form.  Percy and Diebert (2016)
discussed the possibility of some form of rotational variability, but
there were significant problems with that hypothesis.  It is possible,
of course, that different mechanisms act in different stars.  V Hya, for
instance, may be a form of eclipsing binary.

Saio {\it et al.} (2015) proposed that oscillatory convective pulsation
modes might be a possible explanation for the LSPs.  This hypothesis had
some problems, but they might be overcome by a better treatment of
convection; the authors used the standard mixing-length theory of
convection.  Based on the constraints listed above, this hypothesis is
very attractive, but the ``problem of the LSPs" must be considered unsolved
at this point in time.

\medskip

\noindent
{\bf 5. Conclusions}

\smallskip

We have used AAVSO visual and PEP data to determine improved periods and amplitudes
of lower-luminosity, shorter-period pulsating red giants.
Our results extend those of Percy and Deibert (2016) to stars 
with shorter periods -- down to 20-30 days.  As red giants expand, radial
pulsation becomes detectable in early type giants, with periods of about 20 days.  So does the LSP
phenomenon.  With increasing period, the pulsation and LSP amplitudes
both increase, and are approximately equal on average.  When the period is longer than about 100 days, the
LSP amplitude levels off at 0.1 to 0.2.  The LSP is about five times
the fundamental pulsation period in these shorter-period stars, as it is
in longer-period ones, and the LSP amplitude rises and falls
on a time scale of about 20 LSPs.  The LSP phenomenon thus extends to
the shortest periods.  Its nature and cause, however, 
are still unknown, but it must be able to operate in stars with shorter
periods, warmer temperatures, and lower luminosities.

\medskip

\noindent
{\bf Acknowledgements}

\smallskip

We thank the AAVSO observers who made the observations on which this project
is based, the AAVSO staff who archived them and made them publicly available, and the developers
of the VSTAR package which we used for analysis.  This paper is based, in
part, on a short summer research project by undergraduate astronomy \& physics student co-author HL. 
We acknowledge and thank the University
of Toronto Work-Study Program for financial support.  
This project made
use of the SIMBAD database, maintained in Strasbourg, France.

\bigskip

\noindent
{\bf References}

\medskip

\noindent
Benn, D. 2013, VSTAR data analysis software (http://www.aavso.org/node/803).

\smallskip

\noindent
Kafka, S. 2016, observations from the AAVSO International Database (https://www.aavso.org/aavso-international-database

\smallskip

\noindent
Kiss, L.L. and Percy, J.R. 2012, {\it J. Amer. Assoc. Var. Star Obs.}, {\bf 40}, 528.

\smallskip

\noindent
Nicholls, C.P., Wood, P.R., Cioni, M.-R.L., and Soszy\'{n}ski 2009, {\it Mon. Not. Roy. Astron. Soc.}, {\bf 399}, 2063.

\smallskip

\noindent
Percy, J.R. and Desjardins, A. 1996, {\it Publ. Astron. Soc. Pacific}, {\bf 108}, 847.

\smallskip

\noindent
Percy, J.R. {\it et al.} 1996,, {\it Publ. Astron. Soc. Pacific}, {\bf 108}, 138.

\smallskip

\noindent
Percy, J.R. and Guler, M. 1999, {\it J. Amer. Assoc. Var. Star Obs.}, {\bf 27}, 1.

\smallskip

\noindent
Percy, J.R., Wilson,, J.B. and Henry, G.W. 2001, {\it Publ. Astron. Soc. Pacific}, {\bf 113}, 983.

\smallskip

\noindent
Percy, J.R. and Bakos, A.G. 2003, in {\it The Garrison Festschrift},
ed. R.O. Gray, C.J. Corbally, and A.G.D. Philip, L. Davis Press,
Schenectady NY.

\smallskip

\noindent
Percy, J.R. {\it et al.} 2008, {\it Publ. Astron. Soc. Pacific}, {\bf 120}, 523.

\smallskip

\noindent
Percy, J.R. and Abachi, R. 2013, {\it J. Amer. Assoc. Var. Star Obs.}, {\bf 41}, 193.

\smallskip

\noindent
Percy, J.R. 2015, {\it J. Amer. Assoc. Var. Star Obs.}, {\bf 43}, 223.

\smallskip

\noindent
Saio, H., Wood, P.R., Takayama, M., and Ita, Y. 2015, {\it Mon. Not. Roy. Astron. Soc.}, {\bf 452}, 3863.

\smallskip

\begin{table}\small
\caption{Shorter-Period Stars in the AAVSO LPV Binocular Program}
\begin{tabular}{rrrrrr}
\hline
Star & PP (d) & LSP (d) & LSP/PP & PP Amp & LSP amp \\
\hline
$\theta$ Aps & 108.8 & 1023 & 9.4 & 0.14 & 0.11 \\
RT Cnc & 89.3 & 371: & 4.1 & 0.06 & 0.10 \\
TU CVn & 44.8 & 363: & 8.1 & 0.03 & 0.03 \\
V465 Cas & 97.2 & 895 & 9.3 & 0.07 & 0.18 \\
T Cen & 90.5 & -- & -- & 0.54 & -- \\
SS Cep & 100.3 & 958 & 9.5 & 0.06 & 0.09 \\
RR CrB & 55.5 & 630 & 11.3 & 0.04 & 0.05 \\
AF Cyg &  94 & 1439 & 15.3 & 0.13 & 0.08 \\
V1070 Cyg & 62.4 & 640 & 10.3 & 0.04 & 0.05 \\
U Del & 118 & 1166 & 9.9 & 0.05 & 0.21 \\
CT Del & 80.7 & 372: & 4.6 & 0.06 & 0.12 \\
TX Dra & 77.5 & 712 & 9.2 & 0.09 & 0.15 \\
Z Eri & 118.4 & 725 & 6.2 & 0.05 & 0.11 \\ 
RR Eri & 92.6 & 366: & 3.9 & 0.06 & 0.10 \\
g Her & 87.6 & 877 & 10.0 & 0.03 & 0.16 \\
X Her & 176.0 & 667 & 3.8 & 0.06 & 0.08 \\
UW Her & 106.8 & 985 & 9.3 & 0.09 & 0.07 \\
IQ Her & 76.1 & 624 & 8.2 & 0.05 & 0.12 \\
RX Lep & 100.7 & 572 & 5.6 & 0.06 & 0.07 \\
Y Lyn & 133.2 & 1257 & 9.5 & 0.08 & 0.33 \\
SV Lyn & 67.6 & 545 & 8.1 & 0.06 & 0.08 \\
XY Lyr & 121.4 & 1235 & 10.2 & 0.04 & 0.04 \\
GO Peg & 74.9 & 473 & 6.3 & 0.05 & 0.07 \\
TV Psc & 55.1 & 403 & 7.4 & 0.03 & 0.04 \\
$\tau$4 Ser & 111.1 & 1159 & 10.5 & 0.04 & 0.11 \\
W Tri & 105.5 & 765 & 7.3 & 0.05 & 0.08 \\
ST UMa & 90.2 & 625 & 6.9 & 0.04 & 0.07 \\
TV UMa & 53.8 & 653 & 12.2 & 0.03 & 0.06 \\
V UMi & 72.9 & 759 & 10.5 & 0.12 & 0.08 \\
FP Vir & 62.8 & 384 & 6.1 & 0.08 & 0.11 \\
\hline
\end{tabular}
\end{table}

\begin{table}\small
\caption{Red Giants in the AAVSO PEP Program}
\begin{tabular}{rrrrrr}
\hline
Star & PP(d) & LSP(d) & LSP/PP & PP amp & LSP amp \\
\hline

$\chi$ Aqr & 40.2 & 229 & 5.7 & 0.03 & 0.03 \\
RZ Ari & 56.5 & 507 & 9.0 & 0.05 & 0.05 \\
W Boo & 25 & 360:: & 14.4 & 0.02 & 0.04 \\
RS Cnc & 240.8 & 2050 & 8.5 & 0.19 & 0.12 \\
FZ Cep & 81.8 & 743 & 9.1 & 0.11 & 0.10 \\
FS Com & 55.7 & 680 & 12.2 & 0.06 & 0.06 \\
W Cyg & 132 & (259) & (2) & 0.20 & 0.17 \\
AB Cyg & 69 & 525 & 7.6 & 0.10 & 0.13 \\
V1070 Cyg & 62.8 & 639 & 10.2 & 0.04 & 0.05 \\
V1339 Cyg & 34.1 & -- & -- & 0.04 & -- \\
EU Del & 62.5 & 626 & 10.0 & 0.12 & 0.08 \\
AZ Dra & 44 & 359: & 8.2 & 0.04: & 0.10: \\
Z Eri & 78.5 & 730 & 9.3 & 0.10 & 0.13 \\
RR Eri & 92.5 & 742 & 8.0 & 0.12 & 0.10 \\
$\eta$ Gem & 232 & -- & -- & 0.07 & -- \\
IN Hya & 87.8 & 693: & 7.9 & 0.09 & 0.10 \\
R Lyr & 53 & 380: & 7.2 & 0.04 & 0.04 \\
V533 Oph & 58 & 400 & 6.9 & 0.08 & 0.10 \\
$\rho$ Per & 55 & 723 & 13.1 & 0.05 & 0.06 \\
TV Psc & 55.0 & 400: & 7.3 & 0.03 & 0.04 \\
CE Tau & 105 & 1280 & 12.2 & 0.03 & 0.08 \\
TV UMa & 54 & 640 & 11.9 & 0.06: & 0.07 \\
VW UMa & 66 & 621 & 9.4 & 0.06 & 0.07 \\
VY UMa & 122 & 1180 & 9.7 & 0.05 & 0.06 \\
SW Vir & 155 & 1647 & 10.6 & 0.50 & 0.21 \\
FH Vir & 59.6 & 360: & 6.0 & 0.09 & 0.15: \\
FP Vir & 65 & 375: & 5.8 & 0.07 & 0.10 \\
\hline
\end{tabular}
\end{table}

\begin{table}\small
\caption{Amplitude Variations in the LSPs in Shorter-Period Red Giant Stars}
\begin{tabular}{rrrr}
\hline
Star & LSP amp & N cycles & L/LSP \\
\hline
$\theta$ Aps & 0.07-0.17 & 0.75 & 34 \\
RT Cnc & 0.05-0.15 & 4 & 8 \\
TU CVn & 0.02-0.06 & 2 & 50 \\
SS Cep & 0.07-0.16 & 1.5 & 26 \\
AF Cyg & 0.05-0.11 & 1.5 & 21 \\
V1070 Cyg & 0.04-0.06 & 1.5 & 16 \\
U Del & 0.14-0.26 & 0.5 & 21 \\
CT Del & 0.02-0.18 & 1.7 & 20 \\
Z Eri & 0.08-0.19 & 2 & 21 \\ 
g Her & 0.08-0.22 & 0.5 & 46 \\
X Her & 0.04-0.25 & 1 & 60 \\
RX Lep & 0.05-0.13 & 2 & 18 \\
Y Lyn &  0.15-0.40 & 0.5 & 32 \\
SV Lyn & 0.03-0.10 & 2 & 18 \\
XY Lyr & 0.03-0.07 & 1 & 16 \\
TV Psc & 0.02-0.10 & 1.25 & 40 \\
$\tau$4 Ser & 0.07-0.12 & 1.25 & 17 \\
W Tri & 0.06-0.10 & 1.75 & 20 \\
ST UMa & 0.03-0.12 & 2 & 24 \\
TV UMa & 0.03-0.06 & 2 & 15 \\
V UMi & 0.03-0.20 & 0.75 & 26 \\
FP Vir & 0.06-0.17 & 1.5 & 33 \\
\hline
\end{tabular}
\end{table}

\end{document}